\documentclass[preprint,dvips,a4paper,10pt,superscriptaddress,%
               notitlepage]{revtex4-1}

\usepackage{graphicx}

\usepackage[latin1]{inputenc}
\usepackage{ae,aecompl} 
\usepackage[a4paper]{hyperref} 
\hypersetup{
    bookmarks=true,         
    unicode=false,          
    pdftoolbar=true,        
    pdfmenubar=true,        
    pdffitwindow=false,     
    pdfstartview={FitH},    
    pdftitle={Comment on ``Material evidence of a 38 MeV boson''},    
    pdfauthor={J. Bernhard et al.},     
    pdfsubject={Comment on ``Material evidence of a 38 MeV boson''},   
    pdfnewwindow=true,      
    colorlinks=false,       
    linkcolor=red,          
    citecolor=green,        
    filecolor=magenta,      
    urlcolor=cyan           
}

\newcommand{\MeV}{\ensuremath{\textrm{MeV}}}

\begin{document}
\title{Comment on ``Material evidence of a 38 MeV boson''}
\author{J. Bernhard}
\affiliation{Universit\"at Mainz, Institut f\"ur Kernphysik, 55099
  Mainz, Germany}
\author{J.M. Friedrich}
\affiliation{Technische Universit\"at M\"unchen, Physik-Department, 85748
  Garching, Germany}
\author{T. Schl\"uter}
\affiliation{Ludwig-Maximilians-Universit\"at M\"unchen, Department f\"ur
  Physik, 80799 Munich, Germany}
\author{K. Sch\"onning}
\affiliation{CERN, 1211 Geneva 23, Switzerland}
\collaboration{on behalf of the COMPASS collaboration}
\noaffiliation

\maketitle

In a recent preprint~\cite{vanBeveren:2012xz} it was claimed that
preliminary data presented by COMPASS at recent
conferences~\cite{Bernhard:2011ks,Schluter:2011b} confirm the
existence of a resonant state of mass $38\,\MeV$ decaying to two
photons.  This claim was made based on structures observed in
two-photon mass distributions which however were shown only to
demonstrate the purity and mass resolution of the $\pi^0$ and $\eta$
signals.  The additional structures are understood as remnants of
secondary interactions inside the COMPASS spectrometer.  Therefore,
the COMPASS data do not confirm the existence of this state.

\begin{figure}[h]
  \centering
  \includegraphics[width=.4\textwidth]{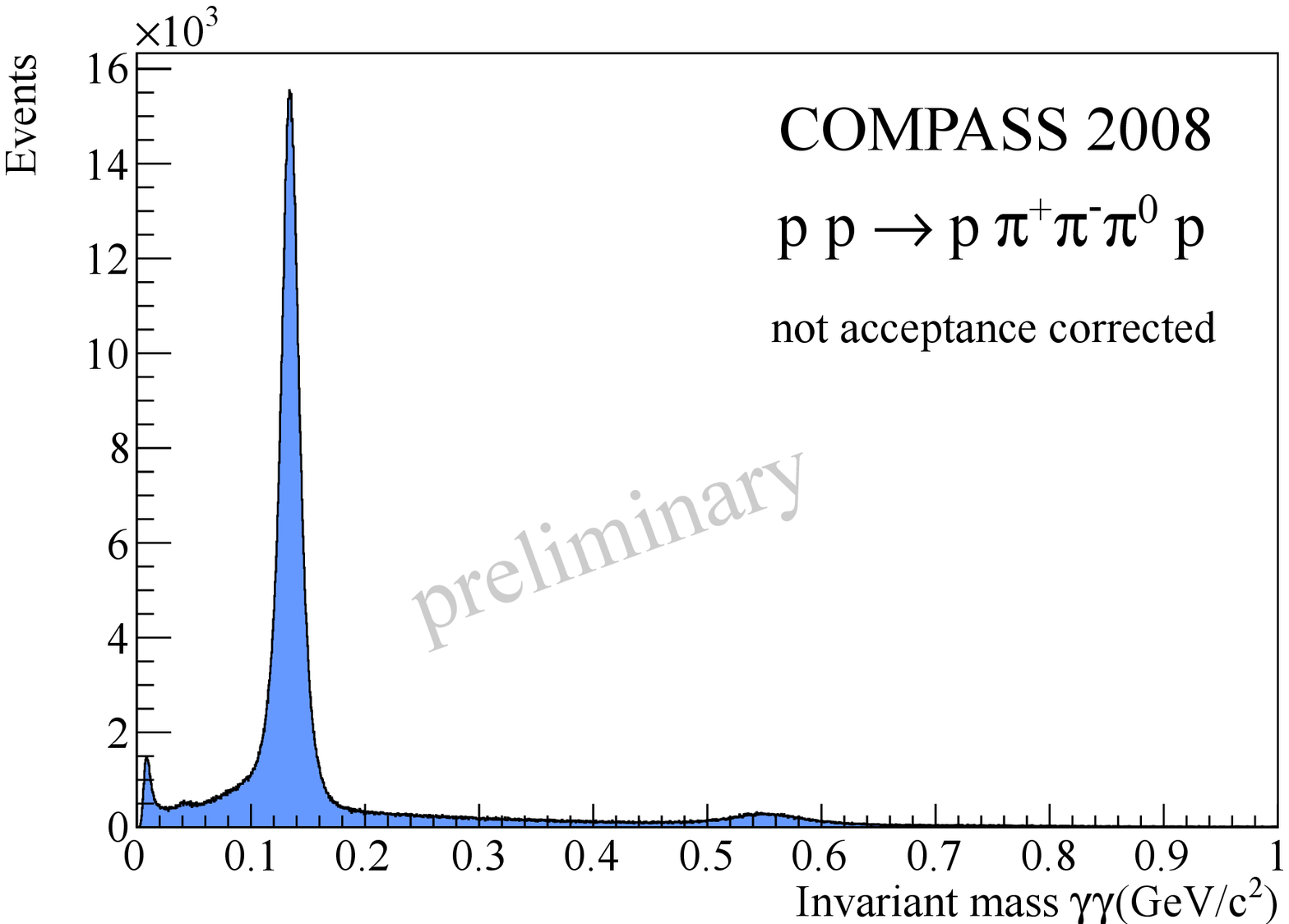}
  \includegraphics[width=.5\textwidth]{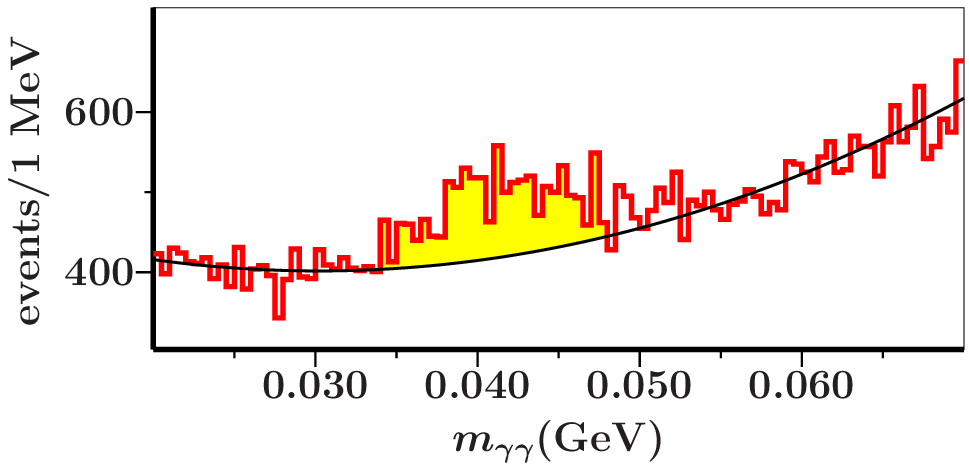}
  \caption{Two-photon mass spectrum from an intermediate step of the
    selection of $pp\to p\omega p$~\cite{Bernhard:2011ks} (left) and the
    plot extracted from this by van Beveren et
    al.~\cite{vanBeveren:2012xz} to support the observation of a light
    boson with mass around $40\,\MeV$.}
  \label{fig:mgg2}
\end{figure}

\begin{figure}[h]
  \centering
  \includegraphics[width=.51\textwidth]{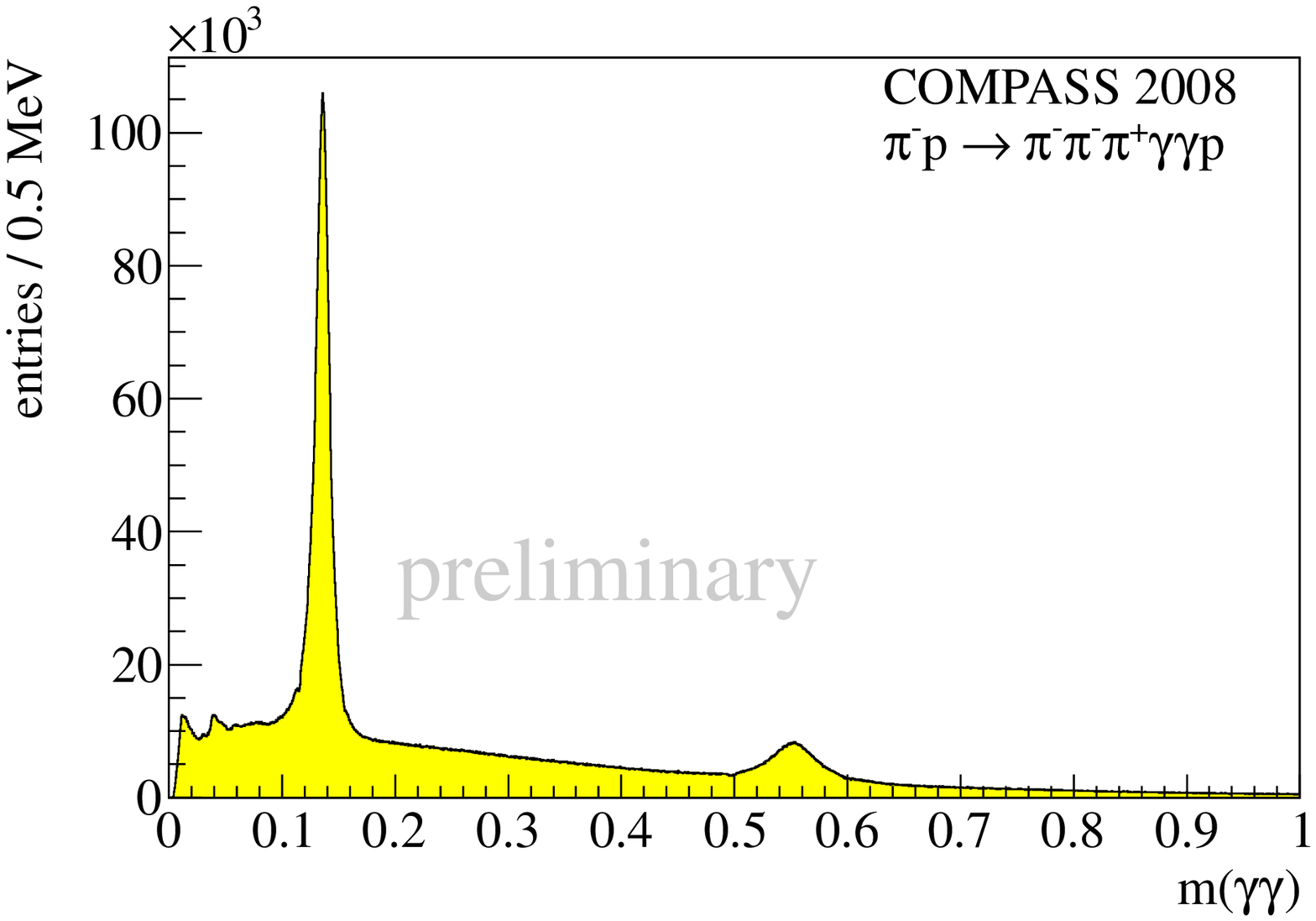}
  \includegraphics[width=.45\textwidth]{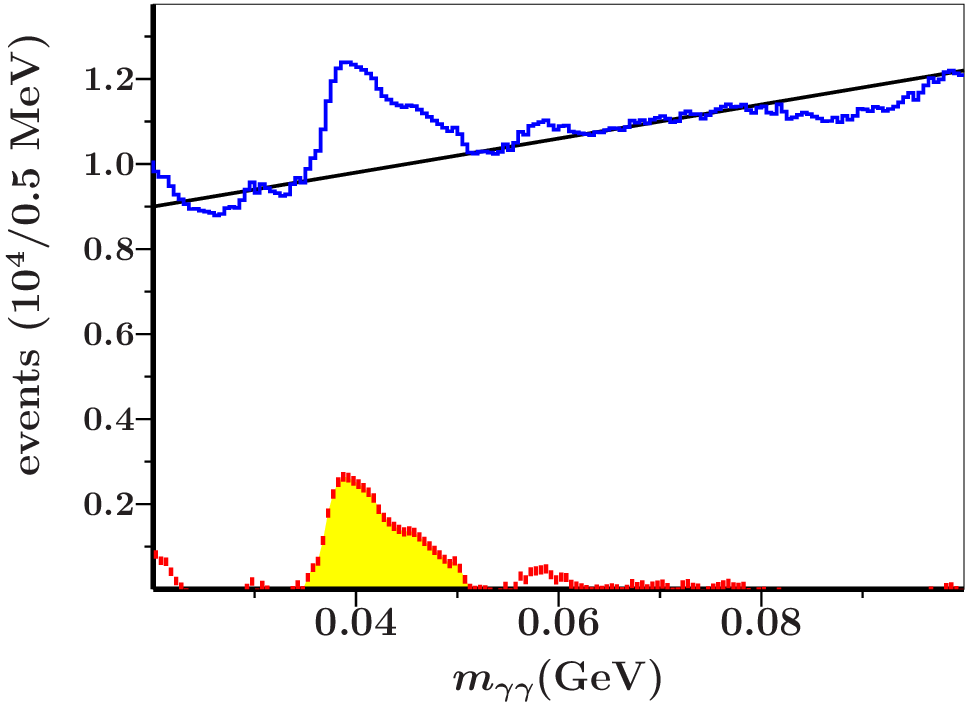}
  \caption{Two-photon mass spectrum from an intermediate step of the
    selection of $\pi^-p\to \pi^-\eta'(\pi^-\pi^+\eta)p$ as shown by
    the COMPASS collaboration~\cite{Schluter:2011b} (left) and the
    plot extracted from this by van Beveren et
    al.~\cite{vanBeveren:2012xz} to support the observation of a light
    boson with mass around $40\,\MeV$.}
  \label{fig:mgg}
\end{figure}

In Figs.~\ref{fig:mgg2} and~\ref{fig:mgg} we show these two-photon mass
spectra next to the plots that were extracted from these and shown
in Ref.~\cite{vanBeveren:2012xz}.

\begin{figure}[h]
  \centering
  \includegraphics[width=.32\textwidth]{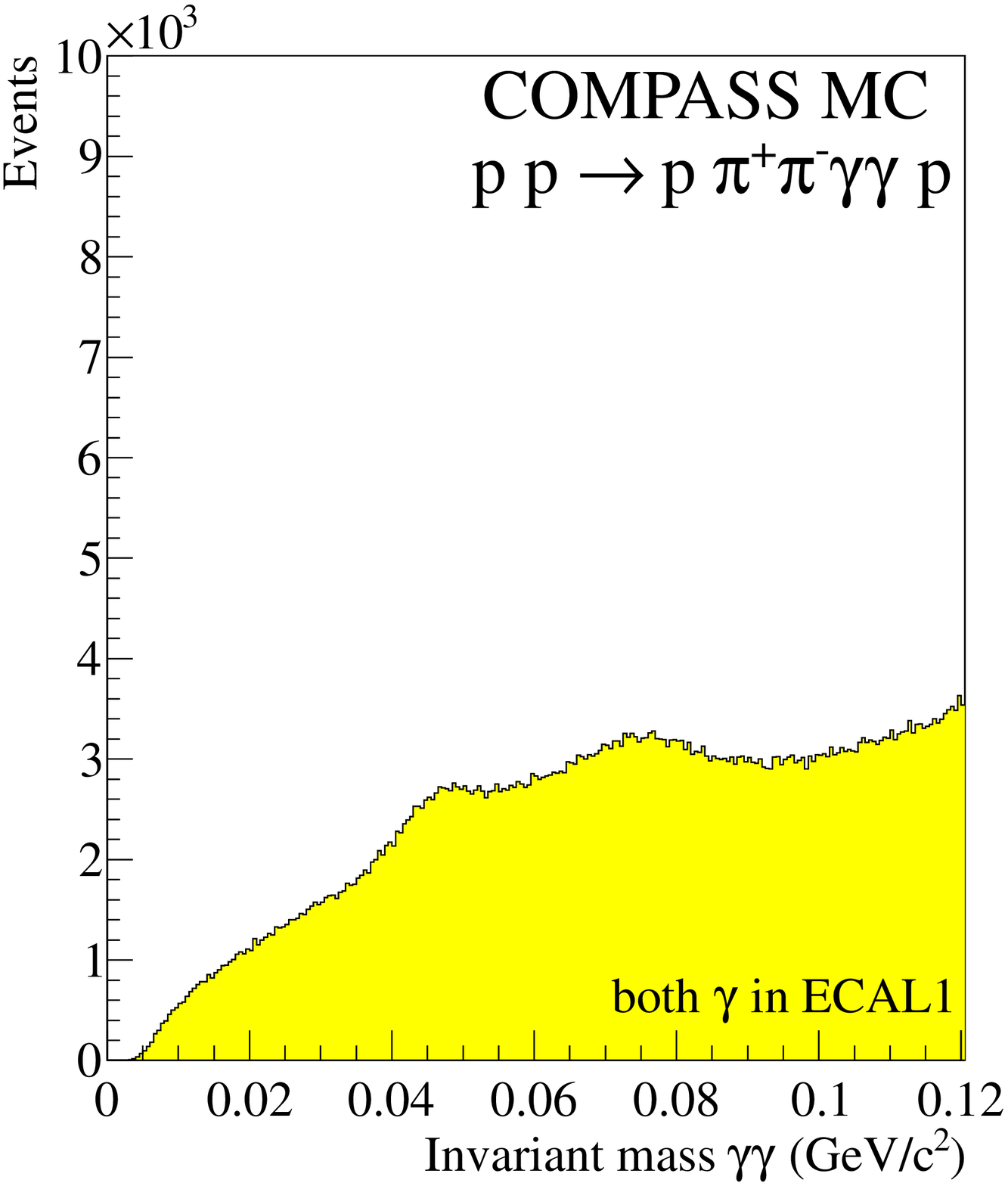}
  \includegraphics[width=.32\textwidth]{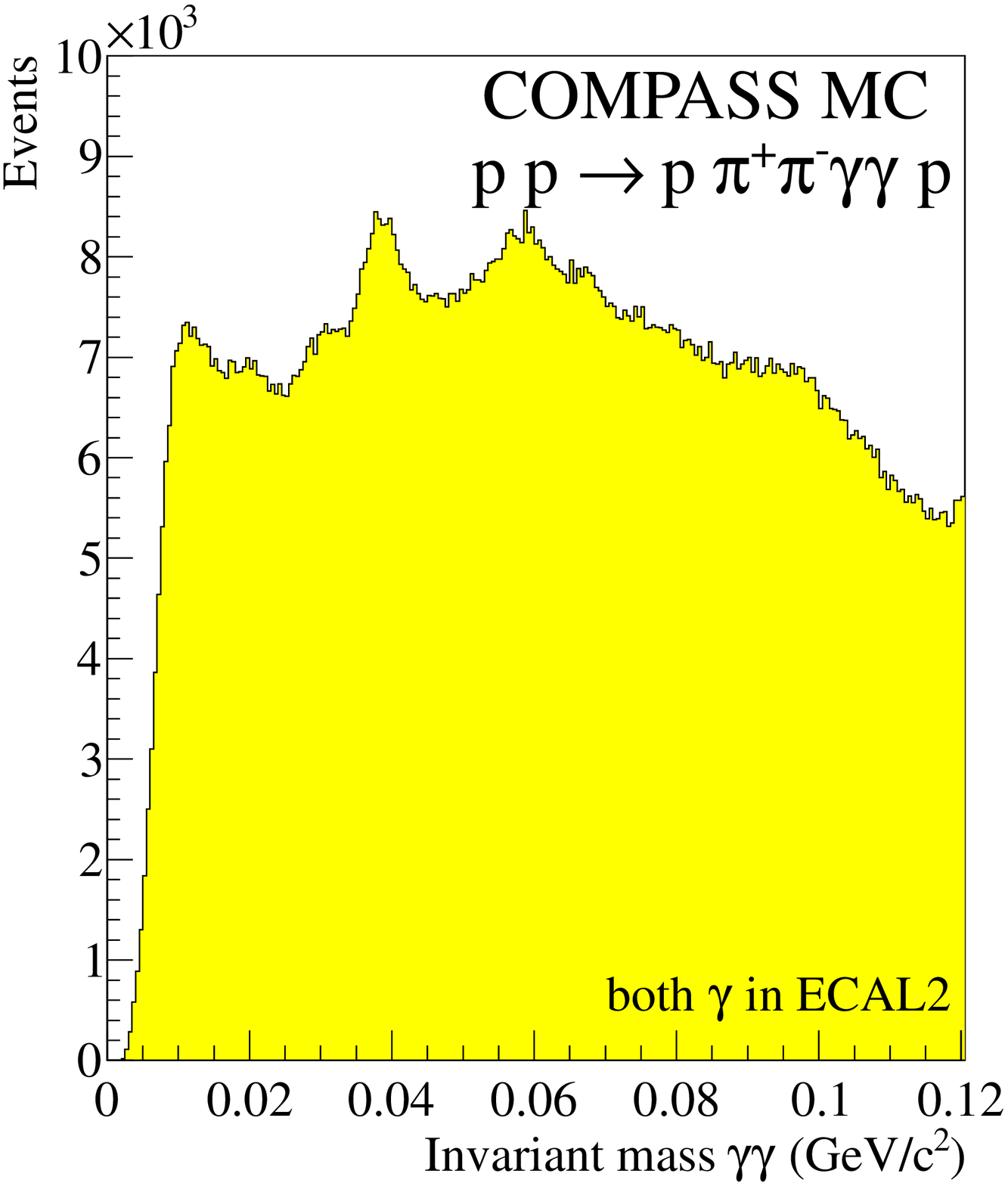}
  \includegraphics[width=.32\textwidth]{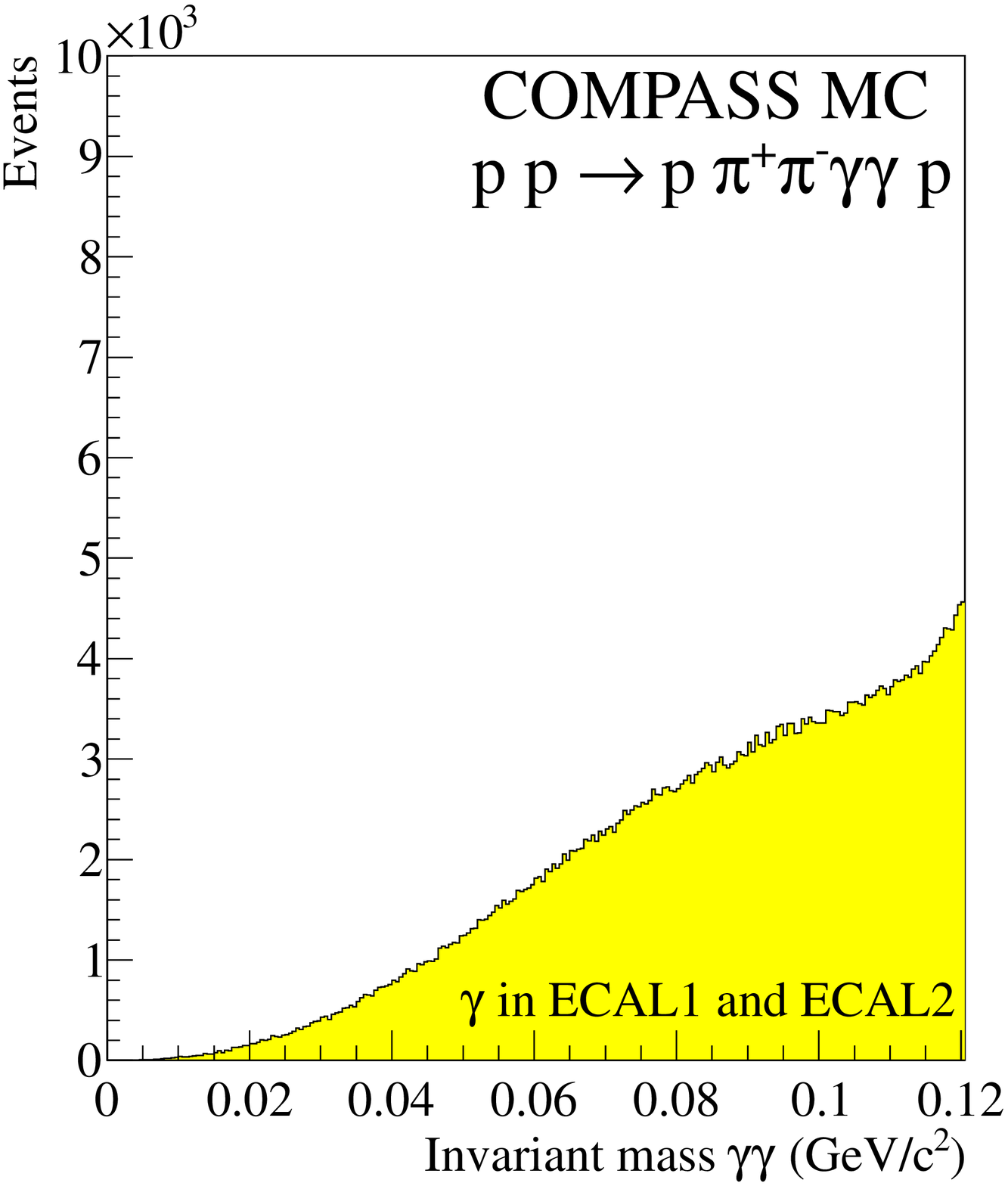}
  \caption{Two-photon mass spectra below the $\pi^0$ peak obtained
    from simulation of in-target interactions $pp\to p\pi^-\pi^+\pi^0
    p$ for the analysis described in Ref.~\cite{Bernhard:2011ks}.  The
    plots show the spectra for the two photons reconstructed in
    different combinations of the calorimeters in the two spectrometer
    stages.  The visible peaks are caused by secondary interactions of
    the outgoing hadrons.}
  \label{fig:mc}
\end{figure}

The structure the authors of Ref.~\cite{vanBeveren:2012xz} focus on is
one from among several such structures observed near the low-mass
edge of the spectrum.  There are several mechanisms how such
structures come about.  Firstly, secondary $\pi^0$ mesons produced in
the detector material downstream of the target lead to
$m_{\gamma\gamma}$ below the nominal $\pi^0$ mass when reconstructed
assuming a target vertex.  Material concentrated in detector groups
leads to peak-like structures.  Likewise, secondary $e^+e^-$ pairs
from photon conversion in the spectrometer material lead to low-mass
structures.  Cuts applied in the reconstruction software lead to
additional structure in the low-mass range.  These artefacts are
reproduced in the Monte Carlo simulation for the reactions under
study, using a complete description of the spectrometer material and
employing the same reconstruction software as for the real data
analysis.  As an illustration of this, we show in Fig.~\ref{fig:mc}
two-photon invariant mass spectra from the Monte Carlo simulations
used in the analysis of $pp\to p\omega p$, $\omega\to \pi^-\pi^+\pi^0$
studied in Ref.~\cite{Bernhard:2011ks}.  We emphasize that no physical
states below the $\pi^0$ mass are included in the MC event generation.

In conclusion, the data do not support any interpretation of these
structures in terms of new resonances.
 
\providecommand{\href}[2]{#2}\begingroup\raggedright\endgroup

\end{document}